\documentclass[a4paper]{jpconf}
\usepackage{graphicx}
\begin{document}
\pagenumbering{arabic}
\pagestyle{plain}
\title{MiniBooNE Oscillation Results 2011}

\author{Zelimir Djurcic}

\address{Argonne National Laboratory, 9700 South Cass Avenue, Argonne, IL 60439,
 USA}

\ead{zdjurcic@hep.anl.gov}

\begin{abstract}
The MiniBooNE neutrino oscillation search experiment at Fermilab has recently updated results from a search 
for $\bar{\nu}_{\mu}\rightarrow\bar{\nu}_e$ oscillations, using a data sample corresponding to $8.58 \times 10^{20}$ 
protons on target in anti-neutrino mode. This high statistics result represent an increase in statistics of 52\% compared to result
published in 2010. 
An excess of 57.7 $\pm$ 28.5 events is observed in the energy range 200 MeV $< E_{\nu} <$ 3000 MeV.
The data favor LSND-like $\bar{\nu}_{\mu}\rightarrow\bar{\nu}_e$ oscillations over a background only hypothesis at 91.1\% confidence level in the energy range 475 $< E_{\nu}< $3000 MeV.
\end{abstract}

\section{Introduction}
\begin{figure}[htb!!!]
\includegraphics[angle=0, width=7.9cm, height=6.5cm]{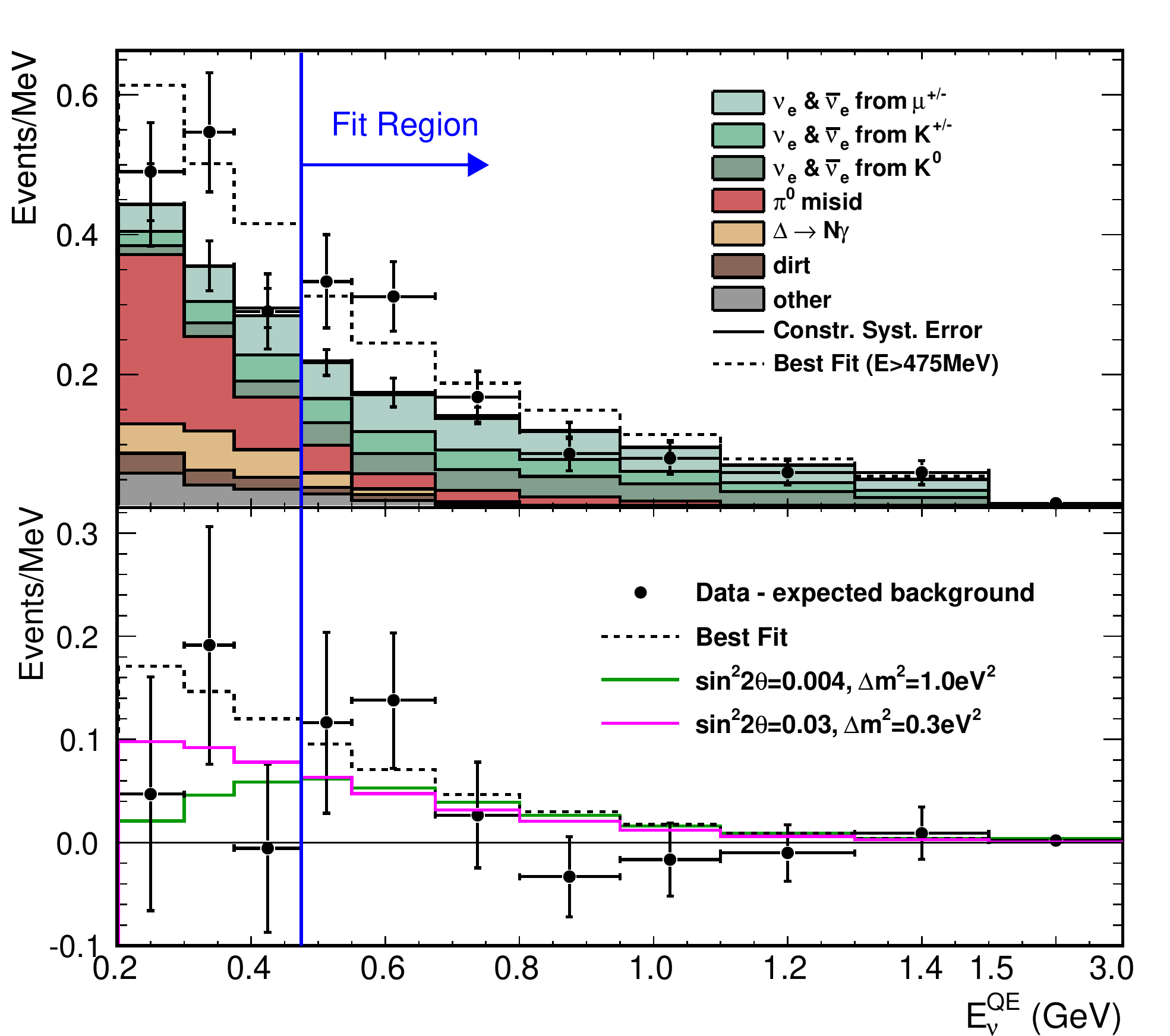} \includegraphics[width=7.9cm, height=6.5cm]{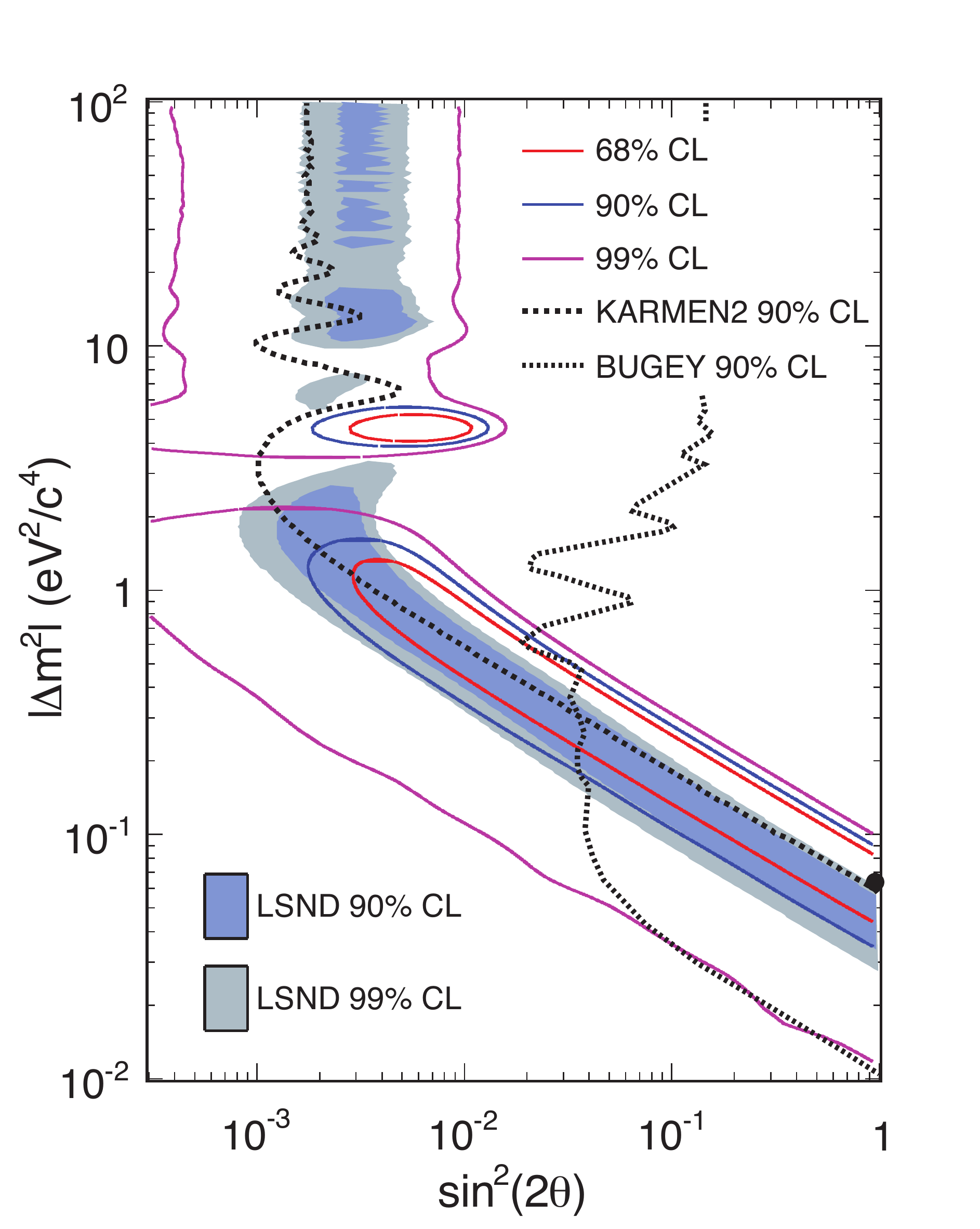}
\caption{Results obtained with  $5.66 \times 10^{20}$ POT exposure. 
Top Left: Reconstructed $E_\nu$ distribution of $\bar{\nu}_e$ CCQE candidates in MiniBooNE anti-neutrino running.  
Bottom Left: The difference between the data and predicted backgrounds as a function of reconstructed neutrino energy. The error bars include both
statistical and systematic components. Also shown in the figure are expectations from the 
best oscillation fit with $E_\nu>475$ MeV, $(\Delta m^2, \sin^2 2 \theta)$ = (0.064 eV$^2$, 0.96) where the fit is extrapolated below 475 MeV
and from other neutrino oscillation parameter sets in the LSND allowed region. 
Right: MiniBooNE 68\%, 90\%, and 99\% C.L. allowed regions for events with 
$E_{\nu} > 475$ MeV within a two neutrino $\bar{\nu}_{\mu}\rightarrow\bar{\nu}_e$ oscillation model, obtained 
with  $5.66 \times 10^{20}$ POT exposure~\cite{mb_osc_antinu_2}.
The shaded areas show the 90\% and 99\% C.L. LSND allowed regions. 
The black dot shows the best fit point.}
\label{limit}
\end{figure}

Motivated by the LSND observation of an excess of observed $\bar{\nu}_e$ events above background prediction in a $\bar{\nu}_{\mu}$ beam~\cite{lsnd}, 
the MiniBooNE experiment was designed to test the neutrino oscillation interpretation of the LSND signal in both neutrino
and anti-neutrino modes. 
The MiniBooNE collaboration has performed a search for $\nu_\mu \rightarrow \nu_e$ oscillations with $6.486 \times 10^{20}$ protons on target (POT), 
the results of which showed no evidence of an excess of $\nu_e$ events for neutrino energies above 475 MeV~\cite{mb_osc_1}.
Despite having observed no evidence for oscillations above 475 MeV, the MiniBooNE $\nu_{\mu}\rightarrow\nu_e$ search observed an excess of 128.8$\pm$43.4 events at low energy, 
between 200-475 MeV~\cite{mb_osc_2}. 
Although the excess is incompatible with LSND-type oscillations within the simple two neutrino oscillation framework, several hypotheses, including sterile neutrino oscillations with CP violation,
anomaly-mediated neutrino-photon coupling,
and many others,
have been proposed that provide a possible explanation for the excess itself~\cite{karagiorgi}.
In some cases, these theories offer the possibility of reconciling the MiniBooNE $\nu_e$ excess with the LSND $\bar{\nu}_e$ excess.
A search in antineutrino mode provides a more 
direct test of the LSND signal, which was observed with antineutrinos.
The MiniBooNE collaboration has published a search for $\bar{\nu}_{\mu}\rightarrow\bar{\nu}_e$ oscillations with $5.66 \times 10^{20}$ POT, 
the results of which showed an evidence of an excess of $\bar{\nu}_e$ events for neutrino energies above 475 MeV~\cite{mb_osc_antinu_2}.
The allowed regions from the fit, shown in Fig.~\ref{limit}, 
are consistent with $\bar{\nu}_{\mu}\rightarrow\bar{\nu}_e$ oscillations 
in the 0.1 to 1 eV$^2$ $\Delta m^2$ range 
and consistent with the allowed region reported by the LSND 
experiment \cite{lsnd}. 
The data show an excess of $43.2 \pm 22.5$ events: 277 electron-like
events have been observed in $200 < E_{\nu} < 3000$ MeV
reconstructed energy range, compared to an expectation 
of $233.8 \pm 15.3(stat) \pm 16.5(syst)$ events~\cite{mb_osc_antinu_2}.
In the energy range $475 < E_\nu< 1250$ MeV, the 
observed $\bar \nu_e$ events, when constrained by the $\bar \nu_\mu$ data 
events, have a $\chi^2/DF = 18.5/6$ and a probability of 
0.5\% for a background-only hypothesis. 

\section{New $\bar \nu_{\mu}\rightarrow\bar{\nu}_{e}$ Result}
\begin{figure}[htb!!!]
\includegraphics[angle=0, width=7.9cm, height=6.5cm]{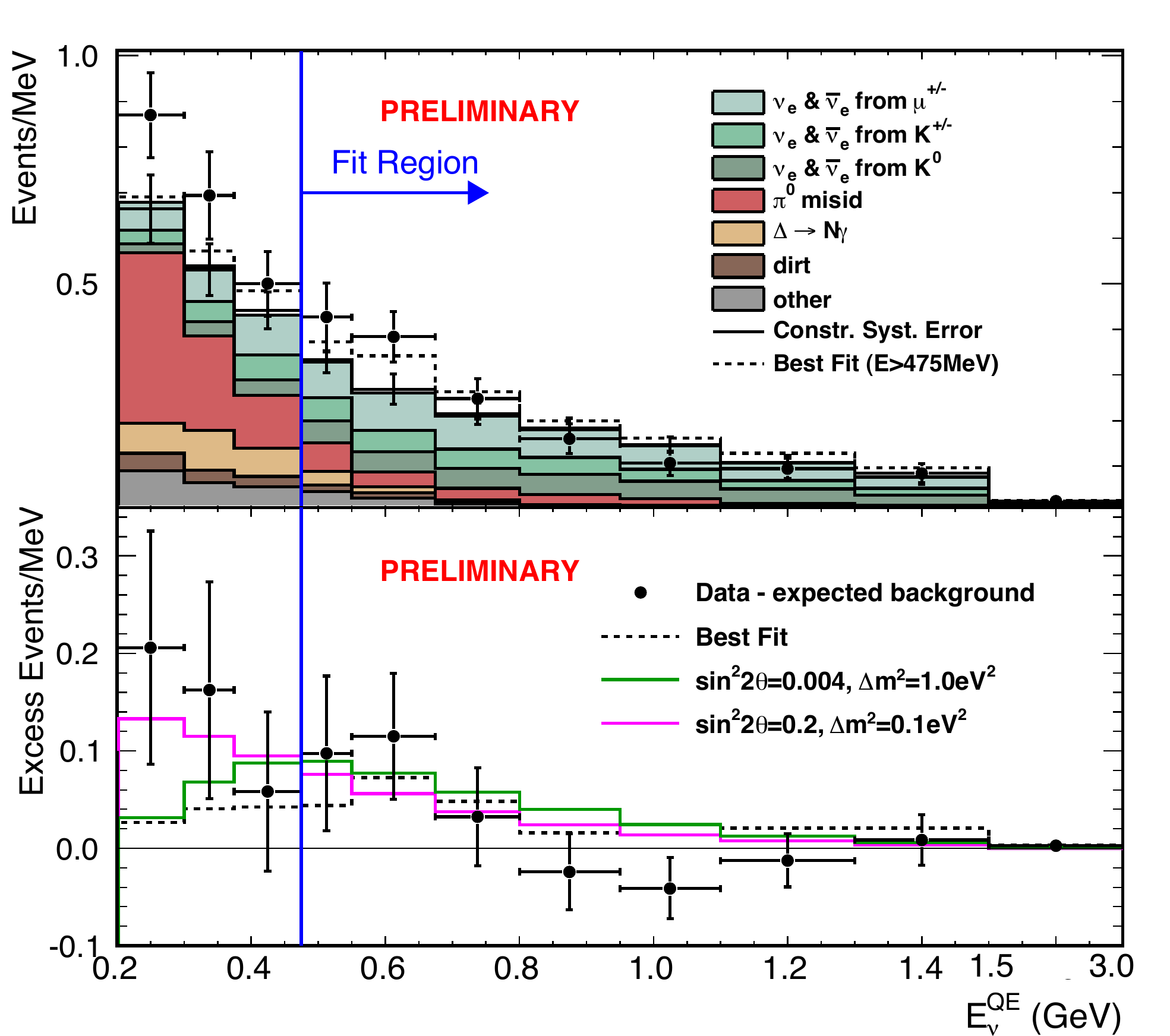}\includegraphics[angle=0, width=7.9cm, height=6.5cm]{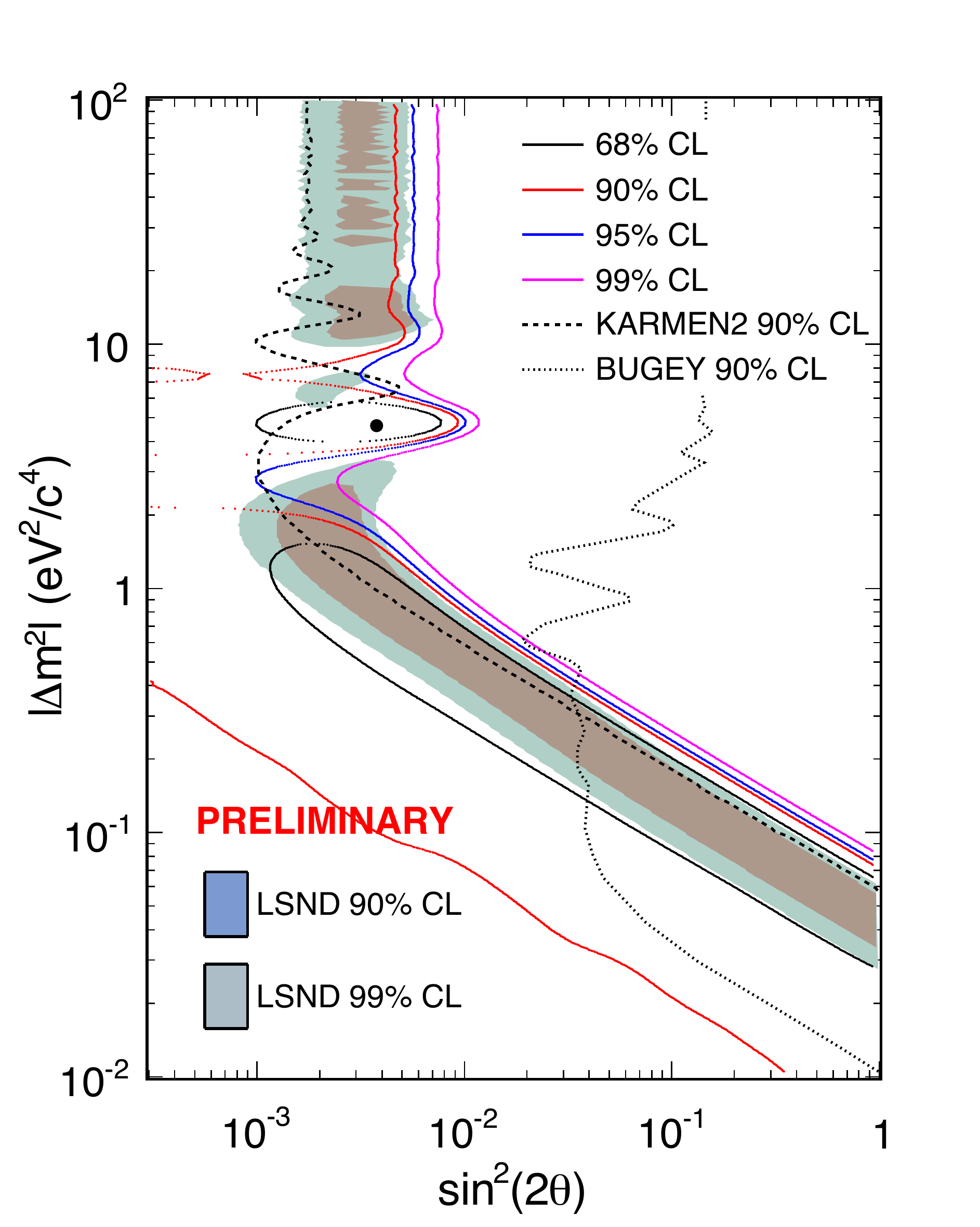}
\caption{Top Left: Reconstructed $E_\nu$ distribution of $\bar{\nu}_e$ CCQE candidates in MiniBooNE anti-neutrino running.  
Bottom Left: The difference between the data and predicted backgrounds as a function of reconstructed neutrino energy. The error bars include both
statistical and systematic components. Also shown in the figure are expectations from the 
best oscillation fit with $E_\nu>475$ MeV, $(\Delta m^2, \sin^2 2 \theta)$ = (4.64 eV$^2$, 0.00337) 
where the fit is extrapolated below 475 MeV
and from other neutrino oscillation parameter sets in the LSND allowed region. 
Right: MiniBooNE 68\%, 90\%, and 99\% C.L. allowed
regions for events with $E_{\nu}  > 475$ MeV within a two
neutrino $\bar \nu_\mu \rightarrow \bar \nu_e$ oscillation model. 
The black dot shows the best fit point.}
\label{excess_new2011}
\end{figure}

\begin{figure}[htb!!!]
\includegraphics[width=7.9cm, height=6.5cm]{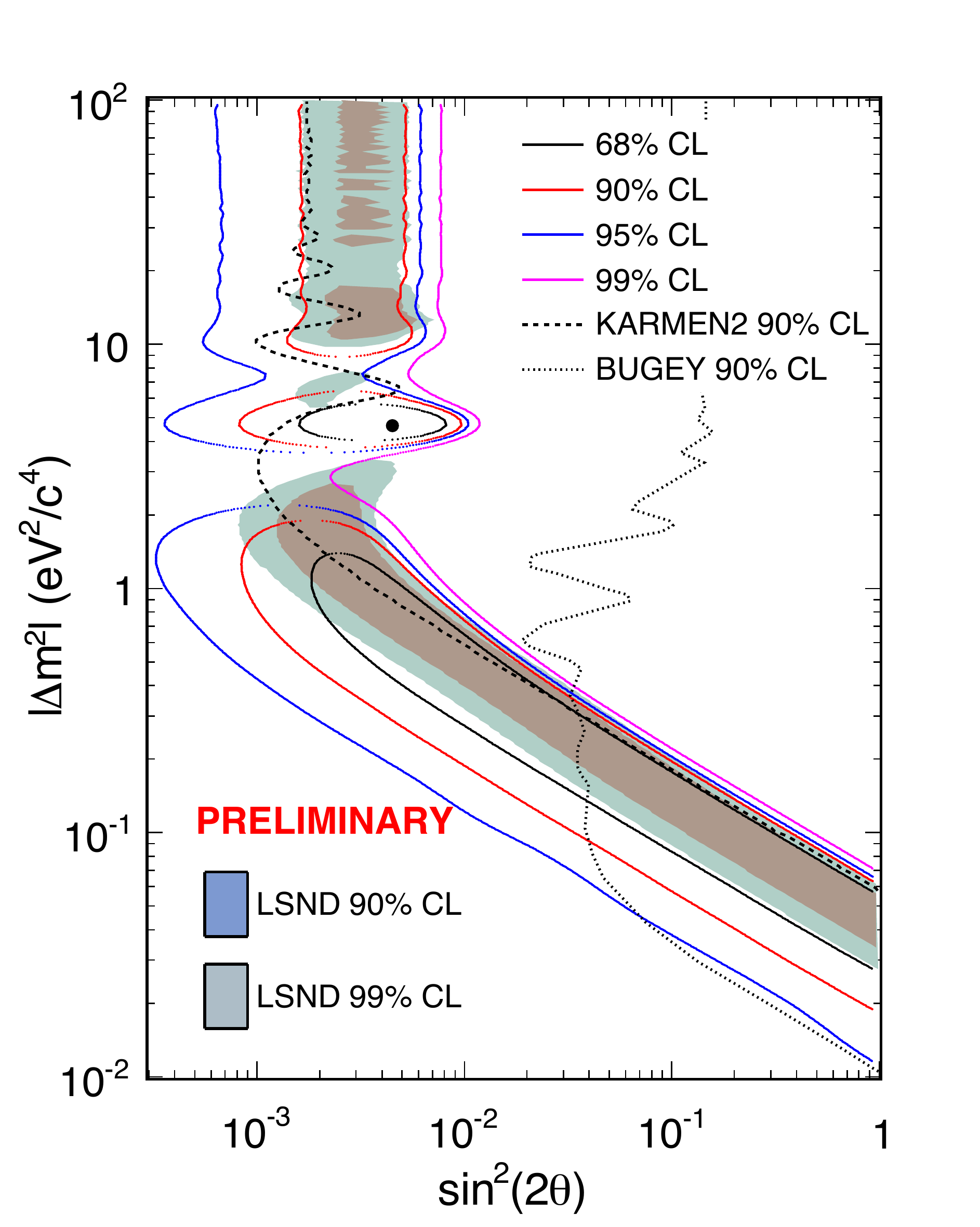}\includegraphics[width=7.9cm, height=6.5cm]{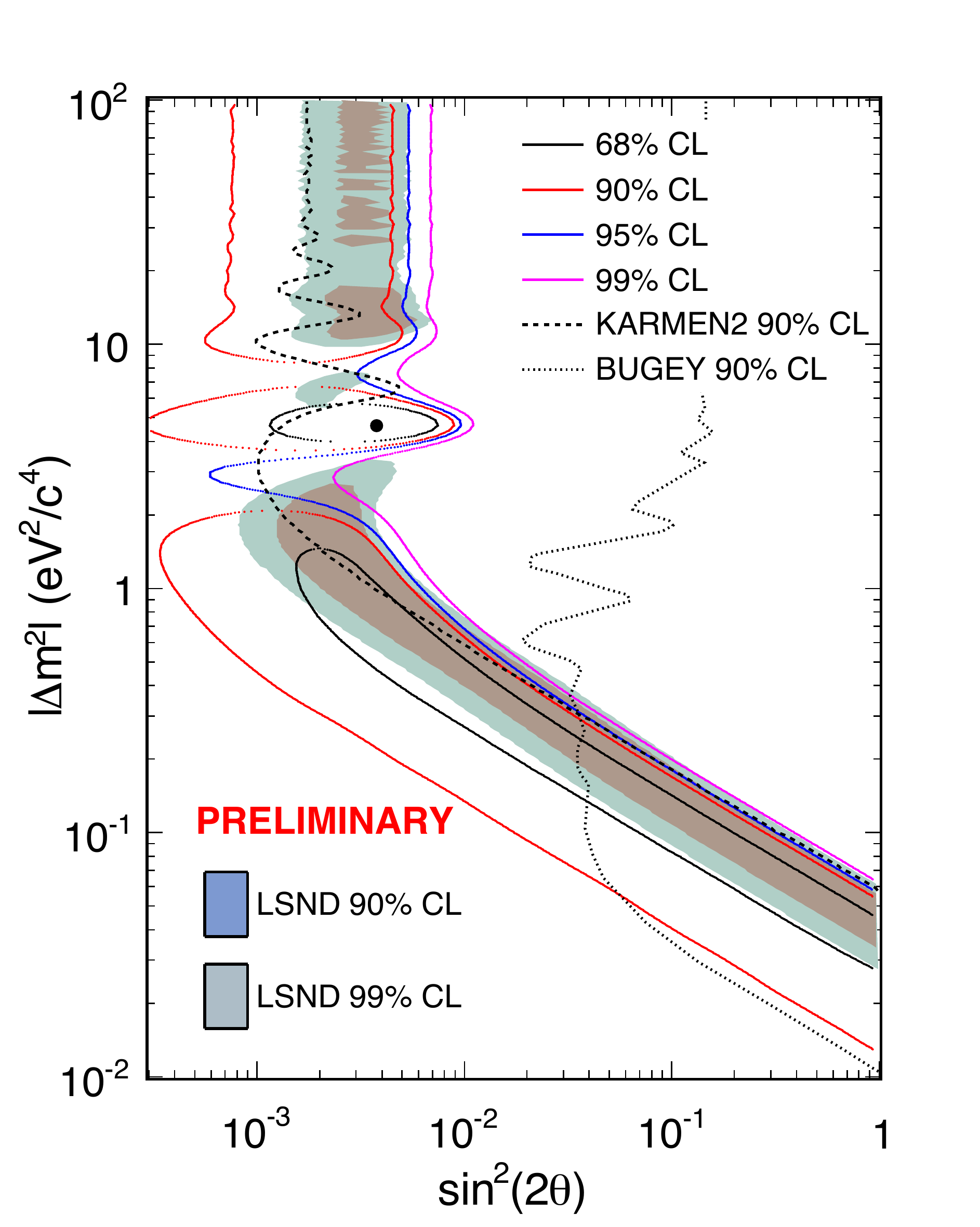}
\caption{Left: MiniBooNE 90\% and 99\% C.L. allowed regions for events with $E_{\nu} > 200$ MeV within a two neutrino
$\bar \nu_\mu \rightarrow \bar \nu_e$ oscillation model. The shaded areas show the 90\% and 99\% C.L. LSND allowed regions.
Right: MiniBooNE 90\% and 99\% C.L. allowed regions for events with $E_{\nu} > 200$ MeV within a two neutrino
$\bar \nu_\mu \rightarrow \bar \nu_e$ oscillation model,
with the subtraction of the expected 17 event
excess in the 200 $< E_{\nu} <$ 475 MeV low-energy region
from the neutrino component of the beam. }
\label{fit200_new2011}
\end{figure}

In this report we describe the latest unpublished results that the MiniBooNE collaboration updated in the search for 
$\bar{\nu}_{\mu}\rightarrow\bar{\nu}_e$ oscillations,
using a data sample corresponding to $8.58 \times 10^{20}$ POT exposure.
The analysis technique used here was already described~\cite{mb_osc_2, mb_osc_antinu_1, mb_osc_antinu_2} 
and assumes only $\bar \nu_\mu \rightarrow \bar \nu_e$ oscillations with no
$\bar{\nu}_{\mu}$ disappearance and no $\nu_{\mu}$ oscillations. 
The signature of $\bar \nu_\mu \rightarrow \bar \nu_e$ oscillations 
is an excess of $\bar \nu_e$-induced charged-current quasi-elastic (CCQE) 
events. 
Fig.~\ref{excess_new2011} (top left) shows the reconstructed $E_{\nu}$ distribution of observed $\bar{\nu}_{e}$ CCQE candidates and background expectation.
\begin{table}[h]
\caption{\label{table_new2011}
The number of data, fitted background, and excess events in the $\bar{\nu}_e$ appearance analysis 
for different $E_{\nu}$ ranges. The uncertainties include both statistical and constrained systematic errors. All known 
systematic errors are included in the systematic error estimate.} 
\begin{center}
{\scriptsize
\lineup
\begin{tabular}{*{4}{l}}
\br                              
\0\0$E_{\nu}$ range [MeV] & Data & Background & Event Excess\cr 
\mr
\0\0200-475& 189  & 150.4 $\pm$ 12.3 $\pm$ 13.9 & 38.6 $\pm$ 18.5 \cr
\0\0475-675& 80  & 58.3 $\pm$ 7.6 $\pm$ 5.5 & 21.7 $\pm$ 9.4 \cr
\0\0475-1250& 168  & 151.7 $\pm$ 12.3 $\pm$ 15.0 & 16.3 $\pm$ 19.4 \cr
\0\0475-3000& 223  & 203.9 $\pm$ 14.3 $\pm$ 20.2 & 19.1 $\pm$ 24.7 \cr
\0\0200-3000& 412  & 354.3 $\pm$ 18.8 $\pm$ 21.4 & 57.7 $\pm$ 28.5 \cr 
\br
\end{tabular}
}
\end{center}
\end{table}
The data show an excess of 57.7 $\pm$ 28.5 events: 412 electron-like
events have been observed in $200 < E_{\nu} < 3000$ MeV
reconstructed energy range, compared to an expectation 
of $354.3 \pm 18.8(stat) \pm 21.4(syst)$ events (see also Table~\ref{table_new2011}).
Fig.~\ref{excess_new2011} (bottom left) shows the event excess as a function of $E_\nu$.
Using a likelihood-ratio technique,
the best MiniBooNE oscillation fit for $475<E_\nu<3000$ MeV occurs at
($\Delta m^2$, $\sin^22\theta$) $=$ (4.64 eV$^2$, 0.00337). 
The energy range $E_{\nu} > 475$ MeV has been chosen
for the fit as this is the energy range MiniBooNE used for
searching for oscillations in neutrino mode. Also, this energy range avoids the
region of the unexplained low-energy excess in neutrino mode~\cite{mb_osc_2}.
The $\chi^2$ for the best-fit point in the
energy range of $475<E_\nu<1250$ MeV
is 4.3 for 4 DF, corresponding to a $\chi^2$-probability of 36\%.
The probability of the background-only fit relative to the best
oscillation fit is 8.9\%. Fig.~\ref{excess_new2011} (right) shows the MiniBooNE
68\%, 90\%, and 99\% C.L. closed contours for 
$\bar{\nu}_{\mu}\rightarrow\bar{\nu}_e$ oscillations in the 
$475<E_\nu<3000$ MeV energy range.
Both the old and updated results are consistent with the LSND oscillation region although the updated result
has a much reduced significance for LSND-like oscillation signal.
The updated results shown an excess of 38.6 $\pm$ 18.5 events in the low energy region 200 $< E_{\nu} <$ 475 MeV, more prominent than in the 
previous result. Part of this excess may be attributed to the neutrino-mode low energy excess~\cite{mb_osc_2}, given 22\% neutrino contribution to the
beam in antineutrino-mode.
With the oscillation fit region extended down to 200 MeV, the MiniBooNE closed contours for $\bar{\nu}_{\mu}\rightarrow\bar{\nu}_e$ 
oscillations are similar as shown in Fig~\ref{fit200_new2011} (left). Fig~\ref{fit200_new2011} (right) shows the fit with the subtraction
of 17 events expected assuming the low energy excess scales with neutrino component of the beam.
The best oscillation fit point without and with this subtraction correspond to $(\Delta m^2, \sin^2 2 \theta)$ = (4.64 eV$^2$, 0.0045) and 
$(\Delta m^2, \sin^2 2 \theta)$ = (4.64 eV$^2$, 0.0037), respectively.
\section{Conclusion and Next Steps}
The MiniBooNE experiment 
observes an excess of 57.7 $\pm$ 28.5 events in the full energy range 200 MeV $< E_{\nu} <$ 3000 MeV for a data sample 
corresponding to $8.58\times10^{20}$ POT. 
The allowed regions from the fit, shown in Fig.~\ref{excess_new2011}, 
are consistent with $\bar{\nu}_{\mu}\rightarrow\bar{\nu}_e$ oscillations 
in the 0.1 to 1 eV$^2$ $\Delta m^2$ range. 
The data favor LSND-like oscillations over a background only hypothesis at 91.1\% 
confidence level.
With new data update excess is indicated at low energy, as with neutrinos.
The MiniBooNE antineutrino results are statistically limited and a larger data sample will be needed to make more definitive statements.
MiniBooNE's requested additional running to reach close to $15 \times 10^{20}$ POT
and significantly increase the current data statistics before anticipated accelerator shutdown at Fermilab expected in Spring 2012.
In the next step a combined $\bar{\nu}_{\mu}\rightarrow\bar{\nu}_e$ and $\nu_{\mu}\rightarrow \nu_e$ analysis will be performed.
In addition, several proposed short-baseline experiments~\cite{new_ideas} will be sensitive to 
$\bar{\nu}_{\mu}\rightarrow\bar{\nu}_e$ oscillations in the 0.1 to 1 eV$^2$ $\Delta m^2$ range.

\subsection{Acknowledgments}
We would like to acknowledge the support of Fermilab, the Department of Energy, and the National Science Foundation.

\end{document}